\documentclass[twocolumn,pra,aps,showpacs,preprintnumbers,amsmath
,amssymb
]{revtex4}

\usepackage{graphicx}
\usepackage{dcolumn}
\usepackage{bm}
\usepackage{subfigure} 
\usepackage{array}
\usepackage{slashbox}
\usepackage{psfrag}
\usepackage{epstopdf}

\begin{document}
\title{Commensurability and hysteretic evolution of vortex configurations in rotating optical lattices}

\author{Daniel S. Goldbaum}
\email{dsg28@cornell.edu}
\author{Erich J. Mueller}\affiliation{
Laboratory of Atomic and Solid State Physics, Cornell University\\
Ithaca, NY 14853
}

\date{\today}

                             
                             \begin{abstract}{
We present a theoretical study of vortices within a harmonically trapped Bose-Einstein condensate in a rotating optical lattice.  Due to the competition between vortex-vortex interactions and pinning to the optical lattice we find a very complicated energy landscape, which leads to hysteretic evolution.  The qualitative structure of the vortex configurations depends on the commensurability between the vortex density and the site density -- with regular lattices when these are commensurate, and concentric rings when they are not.  We model the imaging of these structures by calculating time-of-flight column densities.  As in the absence of the optical lattice, the vortices are much more easily observed in a time-of-flight image than \emph{in-situ}.

   }
\end{abstract}

\maketitle
\newcommand{\adi}{\hat{a}_{i}^{\dagger} }
\newcommand{\ai}{\hat{a}_{i} }
\newcommand{\adj}{\hat{a}_{j}^{\dagger} }
\newcommand{\aj}{\hat{a}_{j} }
\newcommand{\numi}{\hat{n}_{i} }

\section{introduction}
Atomic clouds in rotating optical lattices have garnered a large amount of interest from researchers in the fields of condensed matter physics, atomic physics, and quantum optics. An optical lattice simulates the periodic potential ubiquitous in solid state physics, while rotation probes the superfluid character of these cold atomic gases by driving the formation of quantized vortices. Here we explore the theory of vortices in an optical lattice. Specifically, we investigate the evolution of the vortex configurations that occur in the single-band tight-binding limit as the rotation rate is slowly varied. The energy landscape has a complicated topography that leads to hysteresis. The vortex configurations depend on  commensurability of several different length scales.

A uniform gas of atoms of mass $m$ in an optical lattice rotating with frequency $\Omega$ is characterized by several important scales. Among these are the on-site interaction $U$, the lattice spacing $d$, the magnetic length $\ell=\sqrt{\hbar/m\Omega}$, and the particle density $n$, where $\hbar=h/2\pi$ is Planck's constant.  The behavior of the system changes when these various scales form different commensurate ratios.  There are three well known examples of such commensurability effects, namely when $d^2/\pi\ell^2$ is rational for a two dimensional noninteracting gas, when $n d^D$ is an integer of a non-rotating gas in dimension $D$, or when $\pi n \ell^2$ is rational for a two-dimensional lattice-free gas.  The first example gives the Hofstadter butterfly single-particle spectrum \cite{PhysRevB.14.2239}, the second the superfluid-Mott transition, and the third the fractional quantum Hall effect.  Here we explore how the commensurability between $\ell$ and $d$ plays out in an interacting superfluid,  away from the Mott~\cite{goldbaum:033629,umucalilar:055601, goldbaum-2008}  and fractional quantum Hall limits \cite{bhat:043601, hafezi:023613, PhysRevLett.84.6, palmer:013609, 0953-8984-20-12-123202, mueller:041603, sorensen:086803,palmer}. 

We study the vortex configurations that emerge in a harmonically trapped atomic cloud inside a rotating optical lattice in the single-band tight-binding limit. The resulting phenomenology is rich, as the vortex configurations depend on a number of factors, including: the vortex-vortex interaction, the vortex-pinning potential due to the optical lattice, the finite cloud size, and the past history of the cloud. Fast enough rotation of a uniform superfluid results in the formation of an array of quantized vortex lines of cross-sectional density $n_{vor}$, corresponding to a mean intervortex spacing of $n_{vor}^{-1/2}=\ell/\sqrt{\pi}$. In an infinite system, these vortices arrange themselves in a triangular lattice configuration, but a finite cloud size produces distortions \cite{distorted1,distorted2,distorted3,distorted4}. A co-rotating optical lattice introduces a vortex-pinning potential with minima at the optical lattice potential maxima (between the occupied sites). For commensurate vortex densities, this pinning can cause the lowest energy configuration to switch from
a triangular vortex lattice, to one that shares the geometry of the optical lattice \cite{tung:240402,reijnders:060401,pu:190401,wu:043609,kasamatsu,sato}. 
In practice the vortices are insufficiently mobile to find the true ground state, and one typically sees some metastable configuration, for example with a number of domains separated by defects. We present a realistic simulation of these effects.

We perform calculations in two-dimensions, modeling a harmonically trapped gas of bosons in a rotating square lattice. The two-dimensional case is convenient because we can then concentrate on the interaction between vortex cores in a single plane. This is also an experimentally relevant geometry, as the dimensionality of the system can be  controlled by using an anisotropic harmonic potential, or optical lattice, where the hard trapping direction is along the rotation axis of the optical lattice~\cite{spielman:120402}. A recent experiment~\cite{tung:240402} realized exactly this scenario by placing a rotating mask in the Fourier plane of a laser beam which forms an optical dipole trap.  The mask contained three/four holes, producing a triangular/square lattice in the image plane, where the atoms were trapped.  The lattice spacing was large due to the nature of their optics but can in principle be made  
small enough to explore the single-band tight-binding limit that we investigate.
A similar setup, using a dual-axis acousto-optic deflector, promises to reach this limit in the near future \cite{foot}.

We find hysteresis in our numerical algorithm, reflecting the complicated energy landscape for the vortex configurations, and discuss how similar hysteresis will appear in experiments. This landscape reflects the competition between vortex-vortex interactions and pinning to the optical lattice. In section II we describe our mean-field ansatz and numerical self-consistency routine. In section III we show how vortex configurations evolve from commensurate lattices to incommensurate ring-like structures as the rotation rate is varied. In section IV we present the hysteretic evolution of vortex configurations on spin-up and then spin-down. In section V we simulate the results of time-of-flight imaging of these systems, and in section VI we summarize our results.

 Previous work, focusing on the multi-band weak lattice limit, found vortex structures similar to those we see in our tight binding model 
 \cite{pu:190401,reijnders:060401}, but did not report on how these structures evolved as parameters were adiabatically varied.  Our discussion of the expansion of the rotating cloud initially in the single-band tight-binding regime is also novel.
 
\section{Calculation}

In the reference frame of the rotating optical lattice, our system is described by 
the rotating Bose-Hubbard hamiltonian~\cite{wu:043609,PhysRevLett.81.3108}:
\begin{equation}
\hat{H} 
=-\sum_{\langle i,j \rangle} \left(t_{ij} \adi  \aj  + h.c. \right) 
+  \sum_i \left(\frac{U}{2} \numi \left( \numi - 1 \right) -\mu_i \numi\right)
\label{B}
\end{equation}
where $t_{ij}=t\exp\left[ i \int_{\vec{r}_j}^{\vec{r}_{i}} d\vec{r} \cdot \vec{A}(\vec{r}) \right]$ is the hopping matrix element from site $j$ to site $i$.  The rotation vector potential, which gives rise to the Coriolis effect, is $\vec{A}(\vec{r}) = \left( m/\hbar\right) \left(\vec{\Omega} \times \vec{r} \right)= \pi \nu \left( x \hat{y} -y \hat{x} \right)$, where $\nu$ is the number of circulation quanta per optical-lattice site.  The local chemical potential
 $\mu_i=\mu_0-m \left( \omega^2-\Omega^2 \right) r_i^2 /2$ includes the centripetal potential.  In these expressions,
 $\mu_0$ is the central chemical potential, $\omega$ is the trapping frequency, $\Omega$ is the rotation speed, ${\vec r}_i$ is the position of site $i$,  $m$ is the atomic mass, $\adi$ $\left( \ai \right)$ is a bosonic creation (annihilation) operator,
$\numi = \adi \ai $ is the particle number operator for site $i$,  and $U$ is the particle-particle interaction strength.      
The connection between these parameters and the laser intensities are given by Jaksch et al. \cite{PhysRevLett.81.3108}.
Here, and in the rest of the paper, we use units where the lattice spacing is unity, and we operate exclusively at zero temperature.

Both the superfluid and Mott insulator are well described by 
a spatially inhomogeneous Gutzwiller product \emph{ansatz}~\cite{PhysRevLett.81.3108},
\begin{equation}
\lvert \Psi_{GW} \rangle = \prod_{i=1}^{M} \left( \sum_n f_n^i \lvert n \rangle_i \right) \, ,
\label{C}
\end{equation}  
where $i$ is the site index, $M$ is the total number of sites, $\lvert n \rangle_i$ is the n-particle occupation-number state at site $i$, and $f_n^i$ is the corresponding complex amplitude, with $\sum_n |f_n^i|^2=1$. This ansatz is more general than the more standard mean-field approximation described by the lattice Gross-Pitaevskii equation. In the limit where the latter works well, the two theories agree. The Gutzwiller ansatz has been used extensively to understand experimental results~\cite{folling:060403,GretchenK.Campbell08042006, Greiner:2002lr}, and is well suited for studying the vortex physics that we consider here.

Using equation~\eqref{C} as a variational {\em ansatz}, we minimize the energy with respect to the $\{f_n^i \}$.  We then extract the density $\rho_i=\sum_n n |f_n^i|^2$ and the condensate order parameter $\alpha_i=\langle \hat{a}_i \rangle= \sum_n \sqrt{n} \left(f_{n-1}^{i}\right)^* f_n^i$ at each site. The condensate density $\rho_i^c=\lvert \langle \hat{a}_i \rangle \rvert^2$ is equal to the superfluid density in this model, and is generally not equal to the density.

We use an iterative algorithm to  determine the $\{f_n^i\}$ which (locally) minimize the energy.  We use a square region with $L$ sites per side with hard boundary conditions. We find that we must take 
$L$ much larger than the effective trap radius so that our solutions do not depend on those boundaries.  Typically we use $40\leq L \leq 90$.
We calculate $\langle \hat{H}_{RBH} \rangle$ using equation \eqref{C}.  Minimizing $\langle \hat{H}_{RBH} \rangle$ with respect to $f_n^{i*}$ with the constraint $\sum_n |f_n^i|^2-1=0$ gives $L^2$ nonlinear eigenvalue equations, one for each site,
\begin{widetext}
\begin{equation}
 -t \sum_{k \text{, nn of } j}  \left( \langle \hat{a}_k \rangle  \sqrt{m}  f_{m-1}^j R_{jk} + \langle \hat{a}_k^{\dagger} \rangle \sqrt{m+1} f_{m+1}^j R_{kj} \right) 
+ \left( \frac{U}{2} m^2- \left( \mu\left( r \right)+\frac{U}{2} \right) m + \lambda_j \right) f_m^j = 0 \, ,
\label{M}   
\end{equation}
\end{widetext}
where the sum is over all nearest neighbors of site $j$, $m$ is the particle-number index, $\lambda_j$ is a Lagrange multiplier, and $R_{jk} = \exp$${\left[ i \int_{\vec{r}_k}^{\vec{r}_{j}} d\vec{r} \cdot \vec{A}(\vec{r}) \right]}$, with $i$$=$$\sqrt{-1}$. We iteratively solve these equations: first choosing a trial order-parameter field $\left\{ \alpha_j^{\left( 0 \right)} \right\}$, where $ \alpha_j = \langle \hat{a}_j \rangle $; then updating it by $\alpha_j^{\left( p \right)} = \sum_n \sqrt{n} f_{n-1}^{j*} \left( \left\{ \alpha_j^{\left( p-1 \right)} \right\} \right) f_n^j \left( \left\{ \alpha_j^{\left( p-1 \right)} \right\} \right)$, where $p$ is the iteration index. Similar calculations were performed by Scarola and Das Sarma~\cite{scarola:210403} to analyze the case where the single-particle Mott state is surrounded by a rotating superfluid ring. In the uniform case this algorithm has been used by Wu et. al.~\cite{wu:043609} as well as Goldbaum and Mueller~\cite{goldbaum:033629} and Oktel et. al.~\cite{oktel:045133, umucalilar:055601}.

Since equation (\ref{M}) is highly nonlinear, we find that the solution that this iterative algorithm converges to is sensitive to the initial state we use.  This feature allows us to see the hysteretic effects described in the text.  Experiments will see similar hysteresis, but the precise details will differ from our calculations (for example the critical frequencies for vortex entry and egress will be somewhat modified).

We systematically explore the phase space, varying the parameters in the hamiltonian.  
We simulate clouds with diameter from 30-60 sites, comparable to the sizes studied in experiments~\cite{GretchenK.Campbell08042006, folling:060403}.  For the largest simulations we impose four-fold rotational symmetry, but introduce no constraints in the smaller simulations.

\section{Commensurability and Pinning}

\begin{figure}[tbhp]
\includegraphics[width=\columnwidth]{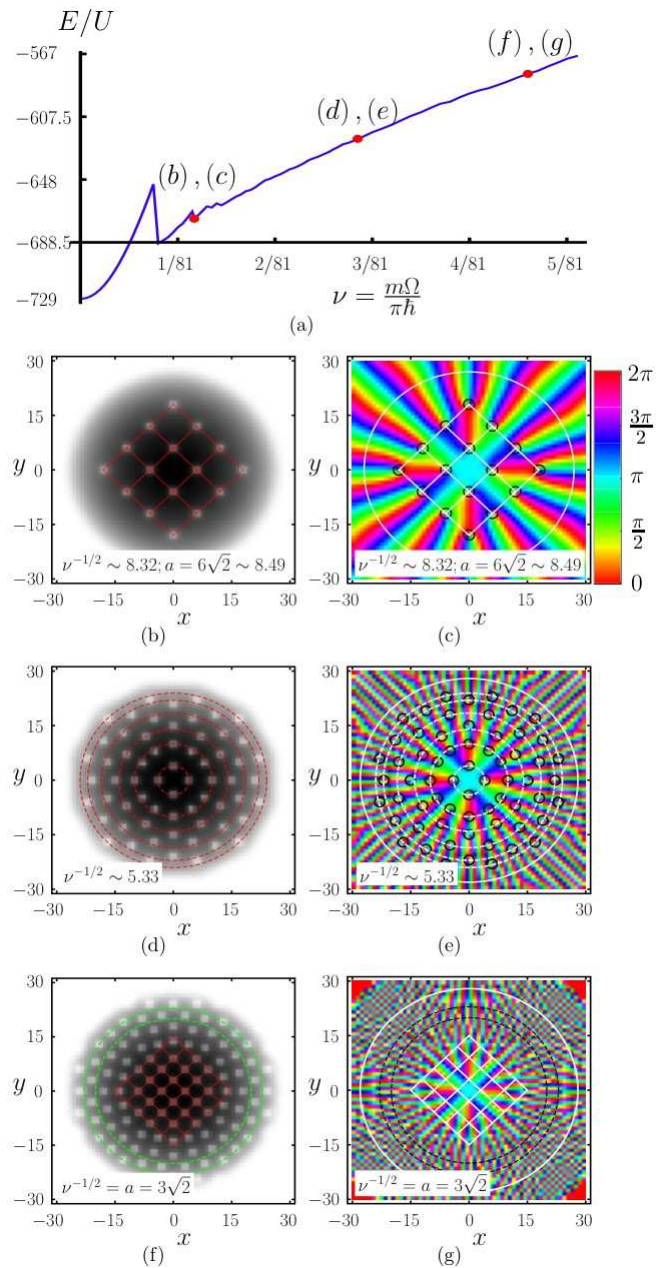}
\caption{\textbf{Adiabatic spin-up} (color-online, one-column).  
Properties of cloud during adiabatic spin-up with parameters
 $\left(t/U=0.2, \, \mu_0/U=0.3, \, R_{TF}=15 \right)$. (a) Energy vs rotation rate. Sharp drops indicate vortex formation. Energy scaled by on-site interaction parameter $U$, and rotation rate quoted as $\nu$, the number of circulation quanta per optical-lattice site. (b), (d), (f) Superfluid density profile at parameters labeled in (a).  Light-to-dark shading corresponds to low-to-high density, and position is measured in lattice spacing. Light spots correspond to vortex cores. Red and green lines are guides to the eye. (c), (e), (g)  Superfluid phase represented by Hue.  Solid white circle denotes edge of cloud.  Dashed lines are guides to the eye.  Black circles denote vortex locations.  In (b), (c) and (f), (g) rotation speed should favor a commensurate square vortex lattice rotated by $\pi/4$ from the optical lattice axes. (d), (e) represents an incommensurate rotation speed.
 }
\label{V-Lattice_Figure_Prep}
\end{figure}
 
  We find a great variety of vortex patterns, including those resembling square vortex lattices.  These are most stable at the rotation rates where they are commensurate with the underlying optical lattice. 
Commensurate Bravais lattices exist when $1/\nu$ is an integer, and commensurate square lattices when $\nu=1/(n^2+m^2)$, for integral $n$ and $m$ \cite{tung:240402,reijnders:060401,pu:190401}.  Which vortex patterns appear in a simulation, or in an experiment~\cite{PhysRevLett.86.4443}, depends on how the system is prepared.
This hysteresis occurs because
 the energy landscape 
 has many deep gorges with near-degenerate energies, separated by high barriers.
 
 To illustrate this energy landscape, 
 we fix $t/U=0.2$ and $\mu_0/U=0.3$, and study how the energy evolves as we vary the rotation speed.
First, starting with the non-rotating ground state, we sequentially increase the rotation speed, using the previous wavefunction as a seed for our iterative algorithm.  We adjust $\omega$ as we increase $\Omega$ so that the cloud size, related to the Thomas-Fermi radius, $R_{TF}=\sqrt{\frac{2 \mu_0}{m\left( \omega^2-\Omega^2 \right)}}$, remains effectively fixed.  
The energy as a function of rotation speed, shown in Figure~\ref{V-Lattice_Figure_Prep}(a), has a series of sharp drops, corresponding to the entry of one or more vortices from outside of the cloud.  At these rotation speeds the system jumps from one local minimum of the energy landscape to another.

 Figure \ref{V-Lattice_Figure_Prep} (b)-(g) shows the superfluid density and phases associated with the vortex patterns found during this {\em adiabatic} increase in rotation speed, where we impose four-fold rotation symmetry about the trap center. 
Subfigures
 (b) and (c) show a regular square vortex lattice 
 seen near the commensurate $\nu=1/(2\times6^2)$.  Subfigures (d) and (e) show the vortex configuration at $\nu\sim1/(2\times 3.76^2)$ which is intermediate between the commensurate values
$\nu=1/(2\times 3^2)$ and  $\nu=1/(2\times 4^2)$.  Rather than forming a square pattern, the vortex configuration appears to be made of concentric rings.  Such ring-like structures also occur for superfluids rotating in hard-walled cylindrical containers \cite{PhysRevB.20.1886}, where boundaries play an important role.  Despite this analogy, it appears that in the harmonic trap these circular configurations are {\em not} a consequence of the circular boundary. When we simulate an elliptical trap, we still find roughly circular vortex configurations.  As one increases $\nu$ towards  $\nu\sim 1/(2\times 3^2)$, a domain containing a square vortex lattice begins to grow in the center of the trap.  As  seen in subfigures (f) and (g), at commensurability the 
 domain only occupies part of the cloud, even though one would expect that a uniform square lattice would be energetically favorable.  The inability of the system to find the expected lowest energy configuration during an adiabatic spin-up is indicative of the complicated energy landscape.

The patterns that we find are largely determined by the symmetry of the instabilities by which vortices enter the system.  For example, even when we do not impose a four-fold symmetry constraint this adiabatic spin-up approach never produces square vortex lattices oriented at an angle other than $\pi/4$ with respect to the optical lattice axes.  On the other hand, we readily produce other commensurate vortex lattices by choosing the appropriate rotation speed and seeding our iterative algorithm with the corresponding phase pattern.      We have verified this approach with square vortex lattices oriented at various angles with respect to the optical lattice, taking $\nu^{-1/2}=\{4,5,6,7,8,9,10\}$,  $(5 \nu)^{-1/2}=\{2,3,4,5,6\}$ and $(10 \nu)^{-1/2}=\{2,3,4\}$.  

\section{Hysteresis}

\begin{figure}[tbhp]
\includegraphics[width=\columnwidth]{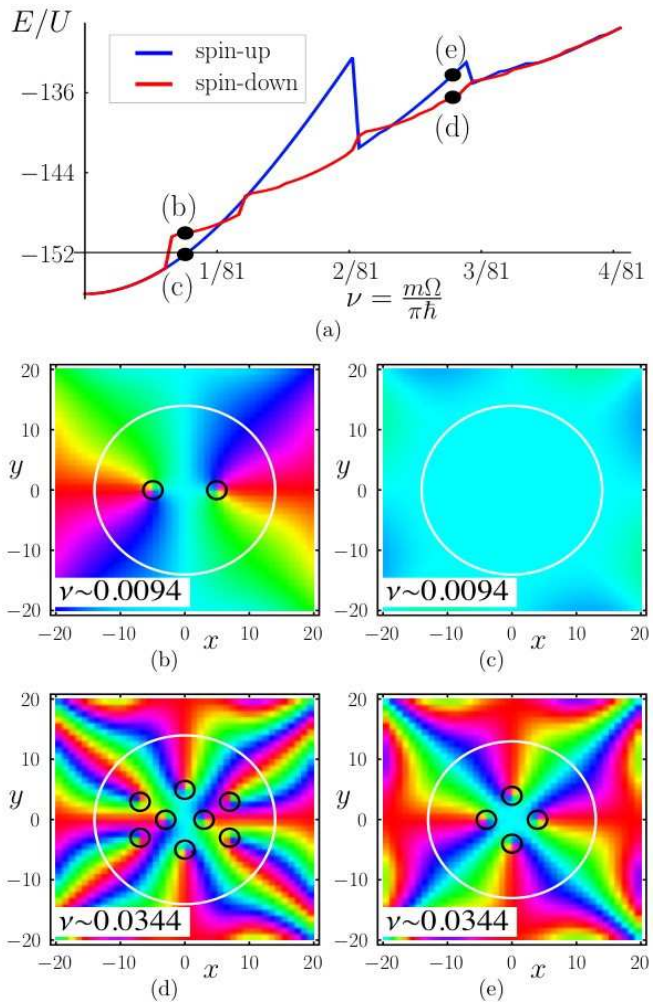}
\caption{\textbf{Hysteresis} (color-online, one-column). (a) Energy  versus rotation rate during increase (blue line) and decrease (red line) of $\nu$.  Energy steps in the blue (red) curve correspond to nucleation (expulsion) of vortices. (b)-(e) Order parameter complex phase for parameters labeled in (a). Black circles are drawn around vortex cores, and white circles indicate the approximate extent of the gas. 
}
\label{Hysteresis}
\end{figure}

We further explore the history dependance of the vortex configurations
by  increasing, then decreasing $\Omega$.
We
do not impose a four-fold symmetry constraint, but take a smaller system with $R_{TF}=7$.
    At any given $\Omega$, the energy shown in figure~\ref{Hysteresis} (a) depends on the system's history.  The increasing(blue)/decreasing(red) rotation curve has sharp energy drops signaling the introduction/ejection of vortices to/from the system.  The energy drops occur at different $\Omega$ for spin-up and spin-down, indicating that the critical rotation speed for a vortex to enter or exit the system is different.   Generically, there are more vortices in the system on spin-down than on spin-up.  Depending on $\Omega$, one may find a lower energy state by increasing  (subfigs. (b) and (c)) or by decreasing (subfigs. (d) and (e)) the rotation rate. As demonstrated by subfigs. (d) and (e), vortex configurations produced during spin-up typically have the four-fold rotational symmetry of the optical lattice, while the vortex configurations calculated during spin-down are more likely to break this symmetry. An experiment will display the same qualitative features, but slightly different vortex configurations.

When $\Omega$ is changed more rapidly ({\em i.e.}, the step-size is increased), we find more symmetry broken states than when we use small steps.  The energy differences between the symmetric and asymmetric states are extremely small, so the energies shown in figure~\ref{Hysteresis} are robust over a large range of sweep rates.

\section{Time-of-flight imaging}
In a previous paper~\cite{goldbaum-2008} we proposed detecting vortices in optical lattice systems through time-of-flight imaging~\cite{Greiner:2002lr,PhysRevLett.84.806}, where at time $\overline{t}=0$ one turns off the lattice and the harmonic trap, letting the cloud expand.  After some fixed time $\overline{t}$ one then produces an absorption image of the cloud using a resonant laser beam.  In a weakly-interacting gas, the density profile is related to the momentum distribution of the gas.  Here we elaborate on this argument, and show how the vortices will be observable in the time-of-flight (TOF) images.
This complements other methods for extracting vortex properties, such as Bragg spectroscopy \cite{vignolo}.  Recently Palmer, Klein, and Jaksch investigated time-of-flight expansion in the fractional quantum Hall limit \cite{palmer:013609}.

We present a simple model where we neglect interactions during the time-of-flight.  This approximation is quite good.  First, the interactions between atoms from different  sites can generically be neglected:  by the time atoms from different sites overlap, the density is so low that they have negligable chance of scattering.  Second, in the single-band tight-binding limit the kinetic energy from the zero-point motion of atoms on a single site should exceed the interaction energy, meaning that the trajectory of the atoms will only be slightly perturbed by the interactions.  If one did include the effects of interactions during the expansion one would see a slight blurring of the interference patterns.  This blurring comes from two effects:
 (1) atoms from sites with higher occupation will be moving faster (the interaction energy is converted into kinetic energy), and (2) the interactions introduce phase shifts which depend on atom number.

Within our approximation, calculating the density of the expanding cloud reduces to a series of single-particle problems.  Taking the initial wavefunction to be given by equation \eqref{C}, after time $\overline{t}$  the wavefunction will be
\begin{equation}
|\psi(\overline{t})\rangle =\prod_{i=1}^{M}  \left(\sum_n f_n^i  \frac{\left[\hat a_i^\dagger(\overline{t})\right]^n}{\sqrt{n!}}\right)\lvert {\rm vac} \rangle,
\end{equation}
where $\hat a_i(0)$ is the operator which annihilates the single-particle state in site $i$ of the lattice.  This operator evolves via the Heisenberg equation of motion, 
$
i \hbar \partial_{\overline{t}} \hat a_j(\overline{t})=\left[ \hat a_j(\overline{t}),\hat{H}_{\rm{free}} \right]
$
where $\hat{H}_{\rm{free}}$ is the Hamiltonian for non-interacting particles.  This is equivalent to evolving the single-particle state annihilated by $\hat a_i(\overline{t})$ via the free Schrodinger equation.

For this analysis we use the notation that $\vec{\bold{r}}$ is a vector in the $x-y$ plane, and $z$ represents the coordinate in the perpendicular direction.
We take the initial (Wannier) state at each site, $\phi_i \left( \vec{\bold{r}},z \right)$, to be gaussian:
\begin{equation}
\phi_i \left( \vec{\bold{r}},z \right)=\frac{1}{\left( \pi \lambda^2 \right)^{1/2}}\frac{1}{\left( \pi \lambda_\perp^2 \right)^{1/4}} \exp{\left[ -\frac{\left(\vec{\bold{r}}-\vec{\bold{r_i}}\right)^2}{2 \lambda^2}-\frac{z^2}{2\lambda_\perp^2} \right]} \, ,
\label{N}
\end{equation}
where $\lambda=\sqrt{\frac{\hbar}{m \omega_{osc}}}$, and $\lambda_\perp=\sqrt{\frac{\hbar}{m \omega_\perp}}$ with $\omega_{osc}$ and $\omega_\perp$ being the small oscillation frequencies for each well. In the geometry we envision, $\omega_\perp\gg \omega_{osc}$.  The wavefunctions at a time $\overline{t}$ after release of the trap are calculated by Fourier transforming $\phi_i \left( \vec{\bold{r}},z \right)$ to momentum space, then time evolving under $\hat{H}_{\rm{free}}$ and finally Fourier transforming back to position space to arrive at $\phi_i \left( \vec{\bold{r}},z,\overline{t} \right)=\phi_i \left( \vec{\bold{r}},\overline{t} \right) f(z,\overline{t})$,  where the only thing we need to know about the transverse wavefunction $f(z,\overline{t})$ is that it is normalized so $\int |f(z,\overline{t})|^2 dz=1$.  The in-plane wavefunction is
\begin{widetext}
\begin{equation}
\phi_i \left( \vec{\bold{r}},\overline{t} \right)= \left( \frac{\lambda^2}{\pi \left( \lambda^2+i \hbar \overline{t}/m\right)^2}\right)^{1/2} \exp{\left[ -\frac{\left(\vec{\bold{r}}-\vec{\bold{r_i}}\right)^2}{2 \left( \lambda^2+i \hbar \overline{t}/m \right)}\right]},
\label{O}
\end{equation}
and the column density of the expanding cloud is then
 \begin{equation}
 n\left( \vec{\bold{r}},\overline{t} \right)=
 \int \langle \psi(\overline{t})| \hat{\psi}^{\dagger}\negmedspace \left( \vec{\bold{r}},z \right)  \hat{\psi} \negmedspace \left( \vec{\bold{r}},z \right) |\psi(\overline{t})\rangle\, dz= \sum_{i=1}^M \left[ n_i - n_{c,i} \right] \lvert \phi_i \left( \vec{\bold{r}},\overline{t} \right) \rvert^2 + \Bigg| \sum_{i=1}^M  \alpha_i \phi_i \left( \vec{\bold{r}},\overline{t} \right) \Bigg|^2 \, ,
 \label{TDep}
 \end{equation}
 \end{widetext}
 where $n_i \, \, \left( n_{c,i} \right)$ is the number of atoms (condensed atoms) initially at site $i$, and $\hat{\psi} \negmedspace \left( \vec{\bold{r}},z \right)$ is the bosonic field operator annihilating an atom at position $(\vec{\bold{r}},z)$. 
 
 In the long-time limit where the expanded cloud is much larger than the initial cloud (i.e.,
  $D_{\overline{t}} = \hbar \overline{t}/m \lambda \gg R_{TF}$), this expression further simplifies, and one has
 \begin{eqnarray}
n(r,\overline{t})&=&
\rho(r,\overline{t})
\left[ 
(N-N_c) 
+ |\Lambda(r,\overline{t})|^2
\right]
\\
\rho(r,\overline{t})&=&\left(\pi D_{\overline{t}}^2 \right)^{-1} e^{-r^2/D_{\overline{t}}^2}
\\
\Lambda(r,\overline{t})&=&\sum_j \alpha_j e^{-i {\bf r \cdot r}_j/D_{\overline{t}} \lambda},
\label{LongTimeLimit}
\end{eqnarray}
where 
$N$ and $N_c$ are the total number of particles and condensed particles, respectively.
The envelope, $\rho(r,\overline{t})$, is a Gaussian, reflecting the Gaussian shape of the Wannier state. 
 The incoherent contribution $(N-N_c) \rho(r,\overline{t})$ has no additional structure. This is a consequence of the Gutzwiller approximation, which neglects short-range correlations.
 Adding these correlations would modify the shape of the background, but it will remain smooth.

The interference term has the structure of the envelope $\rho(r,\overline{t})$ multiplied by the modulus squared of the discrete Fourier transform of the superfluid order parameter.  The discrete Fourier transform  can be constructed by taking the continuous Fourier transform of the product of a square array of delta-functions, and a smooth function which interpolates the superfluid order parameter.  The resulting convolution
produces
 of a series of Bragg peaks, each of which have an identical internal structure which is the Fourier transform of the interpolated superfluid order parameter.  The vortices will be visible in the structure of these peaks.
 
 Vortices in real-space lead to vortices in reciprocal space.  This result is clearest for ``lowest Landau level" vortex lattices~\cite{PhysRevLett.87.060403} which are expressible as an analytic function of $\overline{z}=x+i y$ multiplied by a Gaussian of the form $e^{-|\overline{z}|^2/w^2}$, where $w$ is a length scale which sets the cloud size.  Aside from a scale factor and a rotation, the continuous Fourier transform of such a function is identical to the original.  More generally, the topological charge associated with the total number of vortices is conserved in the expansion process.
 
\begin{figure*}
\includegraphics[width=\textwidth]{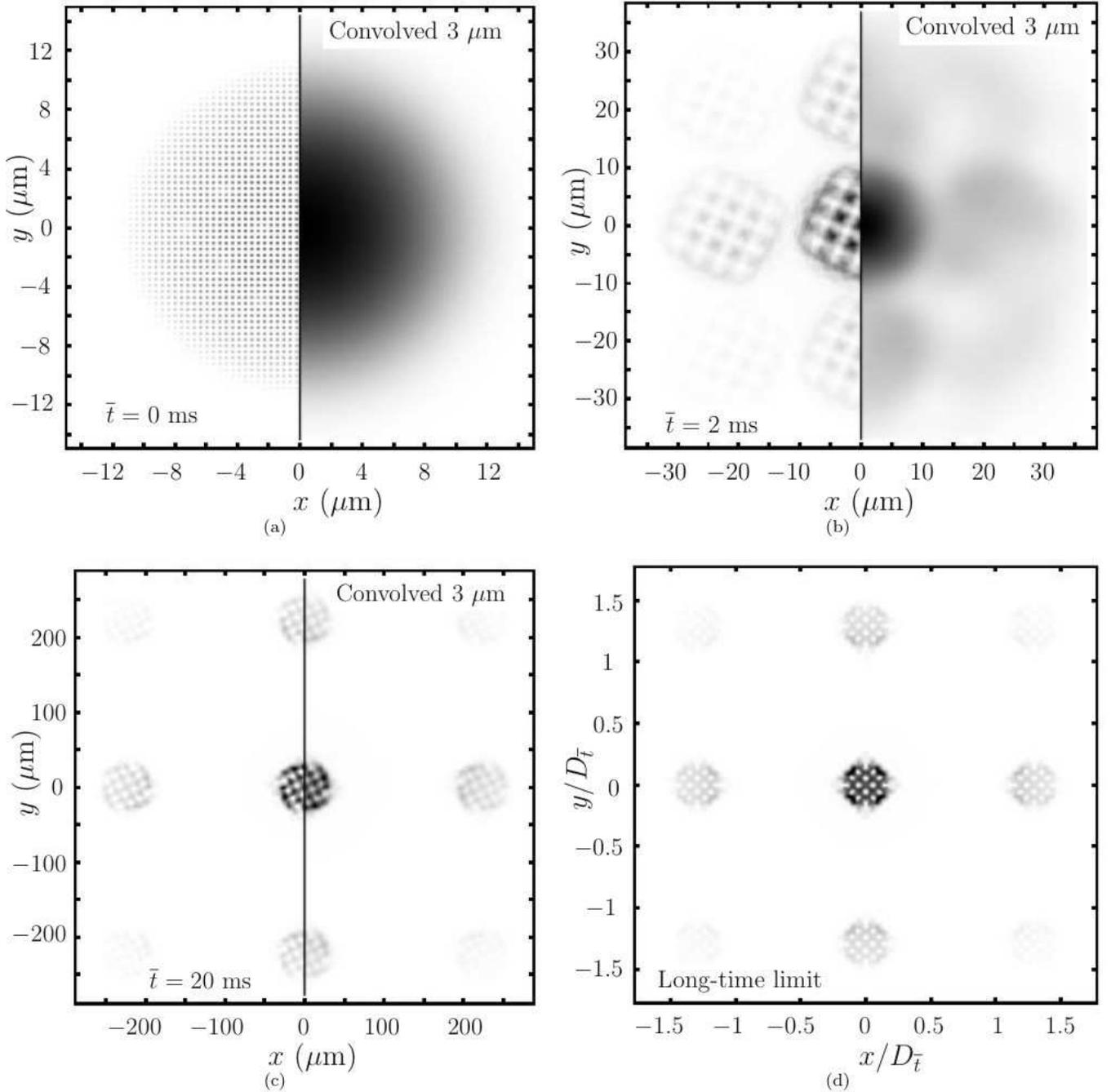}
\caption{\textbf{Simulation of column density seen in time-of flight absorption images}. (Wide). 
Darker shading corresponds to higher column
density. The initial state corresponds to the 
the $4$x$4$ square vortex lattice pictured in subfigures~\ref{V-Lattice_Figure_Prep}(b),(c). (a)-(c) 0, 3, and 20 ms expansion.  Left: column density profile calculated using equation~\eqref{TDep}; Right: column density convolved with a 3~$\mu$m wide Gaussian distribution to represent the finite resolution of a typical imaging system.  (d)  long-time scaling limit, where $D_{\overline{t}} = \hbar \overline{t}/m \lambda \gg R_{TF}$. 
}
\label{TOF_Sim}
\end{figure*}

 Figure~\ref{TOF_Sim} displays simulated expansion images corresponding to the initial square vortex-lattice state shown in subfigures~\ref{V-Lattice_Figure_Prep} (a) and (b), where $t/U=0.2$. In these images,
 lighter colors correspond to smaller column densities.
 Using rubidium-87 atoms on an optical lattice characterized by $d=410$~nm and a hard axis lattice depth of $30$~$E_R$, which are the experimental values in ref.~\cite{spielman:120402}, we find that $V_0=5.7$~$E_R$, which gives $\lambda=84$~nm. Subfigures~\ref{TOF_Sim} (a)-(c) display absorption images: the left-hand side  ($x<0$) is the column density calculated with equation~\eqref{TDep}, while the right-hand side ($x>0$) shows this density convolved with a $3$~$\mu$m wide Gaussian, representing the blurring from typical optics. Subfigure~\ref{TOF_Sim}(d) displays the long-time expansion limit column density calculated using equation~\eqref{LongTimeLimit}, which only depends on the $\{f_n^i\}$'s and the ratio ($\lambda/d$).
   
Subfigure~\ref{TOF_Sim}(a) shows the \emph{in situ} (t=0~ms) column density. At this stage the Wannier functions are tightly peaked about the lattice sites, resulting in a series of sharp density bumps.  The heights of these bumps are slightly modulated due to the square vortex lattice:  near the cores of the vortices there is a slight depletion of the density. Due to the small vortex size, none of this structure is seen once the image is convolved with the Gaussian.  This demonstrates that a typical imaging setup would be unable to resolve the vortices.
Subfigure~\ref{TOF_Sim}(b) shows the abosrption image  after $2$~ms time-of-flight. Several very broad Bragg peaks have developed, each showing a number of low density regions reflecting the square vortex lattice.  Again, the vortex structure is lost upon convolution. Subfigure~\ref{TOF_Sim}(c) shows an absorption image after $20$~ms TOF where, even after convolution, the Fourier transform of the initial vortex pattern is clearly visible in each of the Bragg peaks. In their investigation of atoms in non-rotating optical lattices, Spielman et. al.~\cite{spielman:120402} allowed their atoms to expand for $20.1$~ms before imaging. Subfigure~\ref{TOF_Sim}(d) is a column density calculated in the long-time limit using equation~\eqref{LongTimeLimit}. Aside from an overal scaling, and a slight rotation of the structure within each Bragg peak, the image after 20~ms is almost identical to the image seen in the long-time limit. \\

\section{summary}

We have studied the vortex structures in a harmonically trapped Bose gas in the presence of a rotating optical lattice.  We discussed the hysteretic evolution of the vortex structures as the rotation rate is varied.  This hysteresis is a signature of the complicated energy landscape.  We observed a tendency for the system to find regular lattices configurations when the vortex density is commensurate with the site density.  At incommensurate vortex densities we instead see a circular vortex pattern  which is robust against changing the aspect ratio of the trap.  Finally we analyzed the time-of-flight expansion of one of these condensates.  We find that the vortex patterns are readily observed in the structure of the resulting Bragg peaks.

\section*{Acknowledgements}
We thank Joern Kupferschmidt for 
discussions about the detection method, and for
performing some preliminary time-of-flight simulations. We thank Kaden Hazzard for providing code to calculate the Bose-Hubbard parameters from the underlying continuum model. We also thank Bryan Daniels and Kaden Hazzard for illuminating discussions.
This material is based upon work supported by the National Science Foundation through grant No. PHY-0758104.

\end{document}